\documentclass[twocolumn,showpacs,preprintnumbers,amsmath,amssymb,nofootinbib]{revtex4}
\usepackage{graphicx}
\usepackage{dcolumn}
\usepackage{bm}
\def\jpsi{{J/\psi}}

\def\be{\begin{equation}}
\def\ee{\end{equation}}
\def\bea{\begin{eqnarray}}
\def\eea{\end{eqnarray}}

\def\md{\mathrm{d}}

\def\a{\alpha}

\def\s{\sigma}

\begin{document}
\title{$J/\psi$ polarization in photoproduction up-to the next-to-leading order of QCD}
\author{Chao-Hsi Chang$^{1,2,3}$, Rong Li$^{1,3}$ and Jian-Xiong Wang$^{1,3}$}
\address{$^1$Institute of High Energy Physics, Chinese Academy of Sciences, P.O. Box
918(4), Beijing, 100049, China.\\
$^2$ 
Institute of Theoretical Physics, Chinese Academy of Sciences,
Beijing, 100080, China.\\
$^3$Theoretical Physics Center for Science Facilities, CAS, Beijing, 100049,
China. }


\begin{abstract}
We investigate the $J/\psi$ polarization in photoproduction 
at the Hadron Electron Ring Accelerator(HERA) up to the 
next-to-leading order of QCD. The results show that the transverse
momentum $p_t$ and energy fraction z distributions of $J/\psi$ production do not agree with the observed
ones very well. The theoretical uncertainties for the z distributions of the 
$J/\psi$ polarization parameters with respect to various choices of the renormalization and factorization
scales are too large to give an accurate prediction.
The uncertainties for the $p_t$ distributions of these parameters
are small when $p_t>3GeV$ and 
the obtained $p_t$ distributions can not describe the experimental data even in this region. 
\end{abstract}

\maketitle

The heavy-quarkonium systems $(c\bar{c})$ and $(b\bar{b})$,
especially $J/\psi$ and $\Upsilon$, being flavor hidden, have
received much attention from both experimental and theoretical
sides, since they were discovered.  Their
heavy masses set a ``large" scale that makes the effective field
theory, nonrelativistic quantum chromodynamics
(NRQCD)\cite{Bodwin:1994jh} applicable for their production and decay processes. In
terms of factorization, NRQCD manages the expansion on the strong
coupling constant $\alpha_s$ in the hard part and the $v$ (velocity
between the heavy quark and antiquark in quarkonium) in the soft
part properly. Indeed, NRQCD achieves much success by
taking high Fock state contributions into account. It
reconciles several discrepancies between the theoretical predictions and
experimental data, but there are still some problems. Of the
problems, the discrepancies between the theoretical predictions at leading order (LO) and
experimental data on the polarization of $J/\psi$ and
$\Upsilon$ hadroproduction are outstanding.
Recent reviews on the situation can be found in Ref.~\cite{Brambilla:2004wf}.

Many works~\cite{Kramer:1995nb,Zhang:2005cha,Li:2008ym} indicate that the
next-to-leading-order (NLO) QCD corrections under the NRQCD framework
drastically change the features of LO theoretical
predictions in the heavy-quarkonium production in various cases. Especially, it
is reported in Ref.\cite{Kramer:1995nb} that with the color-singlet 
model (CSM) alone the NLO prediction on $J/\psi$
photoproduction at HERA can explain the
experimental data well. Also with the CSM alone, the NLO results on
the charmonium production moderate, even resolve, the discrepancies
between theoretical predictions and experimental measurements at
$B$ factories \cite{Zhang:2005cha}. Recently,
the NLO QCD corrections on the hadronic production of $J/\psi$ and
$\Upsilon$ in the CSM have also been studied by several
groups\cite{Campbell:2007ws,Gong:2008sn,Artoisenet:2008fc} and  
the results show that the $p_t$ distribution of $J/\psi$ production
is largely enhanced. Furthermore, 
the NLO QCD corrections on the hadronic production of $J/\psi$ in the color-octet
mechanism (COM) have been completed in Ref.~\cite{Gong:2008ft} and  
the results show that the $p_t$ distribution of $J/\psi$ production
is changed slightly, whereas the polarization of the produced quarkonium still is an
open problem. Namely, the results at NLO in the CSM
give a longitudinal polarization contrary to the transverse one
as LO does\cite{Gong:2008sn}, and the NLO results in the COM give 
almost unchanged polarization as the LO ones
\cite{Gong:2008ft}. 
To add the NLO CSM and NLO COM results together, the data for polarization of $J/\psi$ hadroproduction 
at Tevatron cannot be described
properly, although the data for $p_t$ distribution of hadroproduction 
can be fit quite well \cite{Gong:2008sn,Gong:2008ft}.

It may provide $ep$ and $\gamma p$ processes to investigate the production of $J/\psi$ at HERA. 
The transverse momentum $p_t$ distributions of $J/\psi$ production and polarization 
at LO were studied in the CSM a long time 
ago\cite{Berger:1980ni,Baier:1981zz,Korner:1982fk,Korner:1982fm}.
As mentioned above, it was reported in Ref.\cite{Kramer:1995nb} that
with the CSM alone the NLO results can give a proper
description for the photoproduction of $J/\psi$ at HERA, including the energy fraction
distribution $d\sigma/dz$ in the intermediate $z$ region and
transverse momentum $p_t$ distribution, where the energy fraction 
$z$ is defined by $z\equiv \frac{(p_{J/\psi}\cdot
p_p)}{(p_{\gamma}\cdot p_p)}$ and $p_{J/\psi}$, $p_\gamma$ and $p_p$
are the momenta of $J/\psi$, $\gamma$ and proton, respectively. Then, to
investigate the $J/\psi$ inelastic
photoproduction, many works based on NRQCD at LO and NLO in the CSM
and COM followed\cite{Cacciari:1996dg}. In Ref.\cite{Beneke:1998re}
the photoproduction of polarized $J/\psi$ was studied at LO in the CSM
and COM. In Ref.\cite{Saleev:2002gt} the theoretical predictions 
based on the $k_T$ factorization formula were also given at LO.
However, the latest experimental results on the photoproduction of
polarized $J/\psi$\cite{Jungst:2008ip} do not favor the LO
predictions in the CSM and COM, even those obtained with the $k_t$
factorization formula. So, up to now there is no satisfactory
explanation on the measurement results of photoproduction of
polarized $J/\psi$ at HERA. In view of the proper description of the HERA
data on $p_t$ and $z$ distributions by the CSM NLO prediction\cite{Kramer:1995nb}, 
we focus the theoretical predictions on $J/\psi$
polarization with NLO QCD corrections in the CSM in this paper.

To calculate the NLO QCD correction for the photoproduction of polarized $J/\psi$ in the CSM, 
there are direct and resolved processes to be considered. 
The existing LO calculation~\cite{Beneke:1998re}
shows that the contribution from resolved processes is about 2 orders of magnitude smaller 
than that of the direct one for the $p_t$ distribution of $J/\psi$, and only 
in the lower z (z$<$0.2) region is the contribution from resolved processes 
comparable with that from the direct one. Therefore, with the cut condition $z>0.2$ for 
the experimental measurements\cite{Jungst:2008ip,Chekanov:2002at}, 
only the direct processes are investigated in this paper. 
To perform the lengthy analytic evaluation for all the processes, 
the computer program package, Feynman Diagram Calculation (FDC), is used.
FDC was developed and well
tested for many years\cite{Wang:2004du}, and recently the functions
for manipulating one-loop quantum corrections in analytic reduction
and numerical calculations were completed and tested in many
aspects~\cite{fdc:2008}.
\begin{figure*}
\center{
\includegraphics*[scale=0.31]{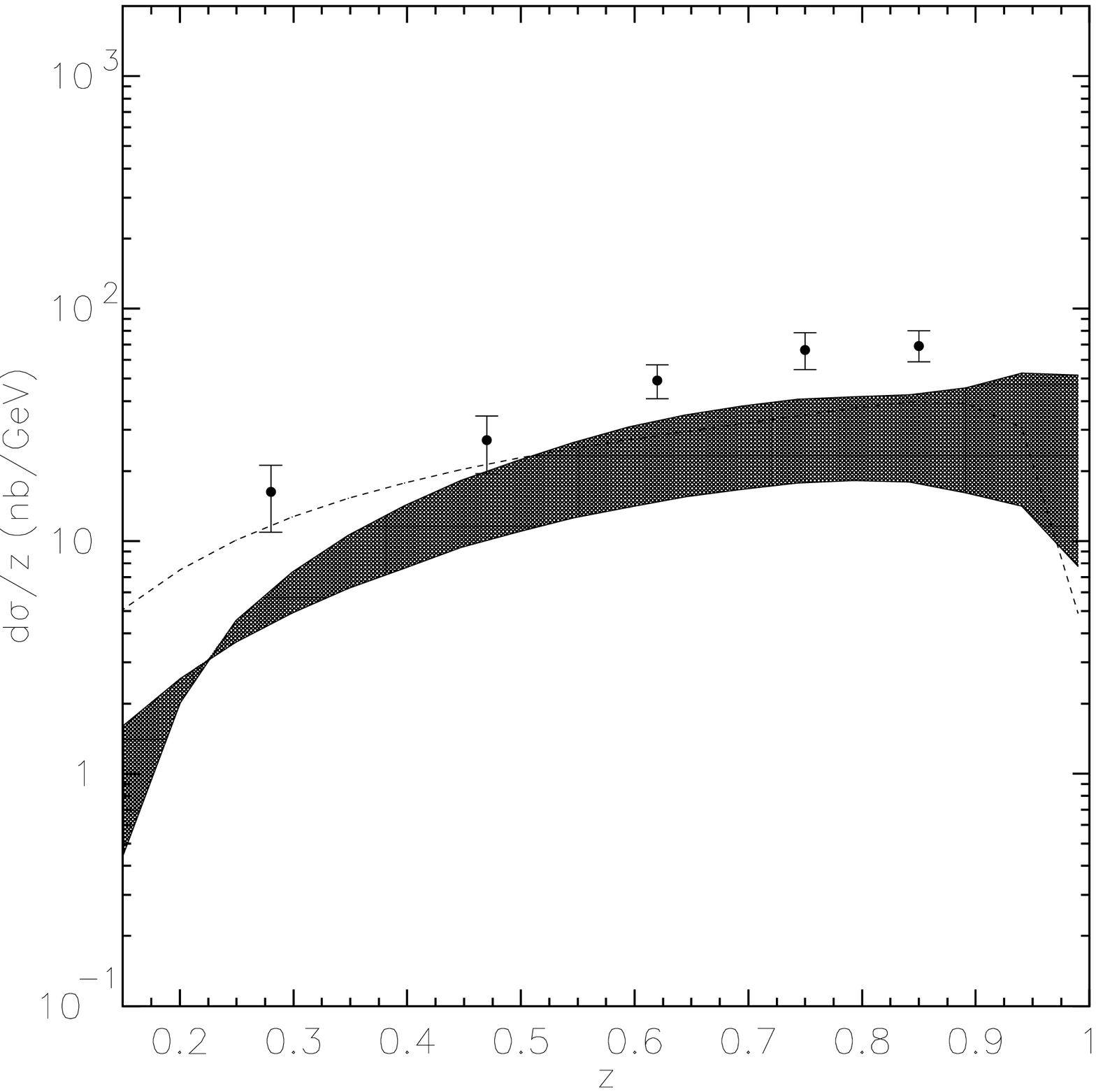}
~~~~~~~~~~~~~~~~~
\includegraphics*[scale=0.31]{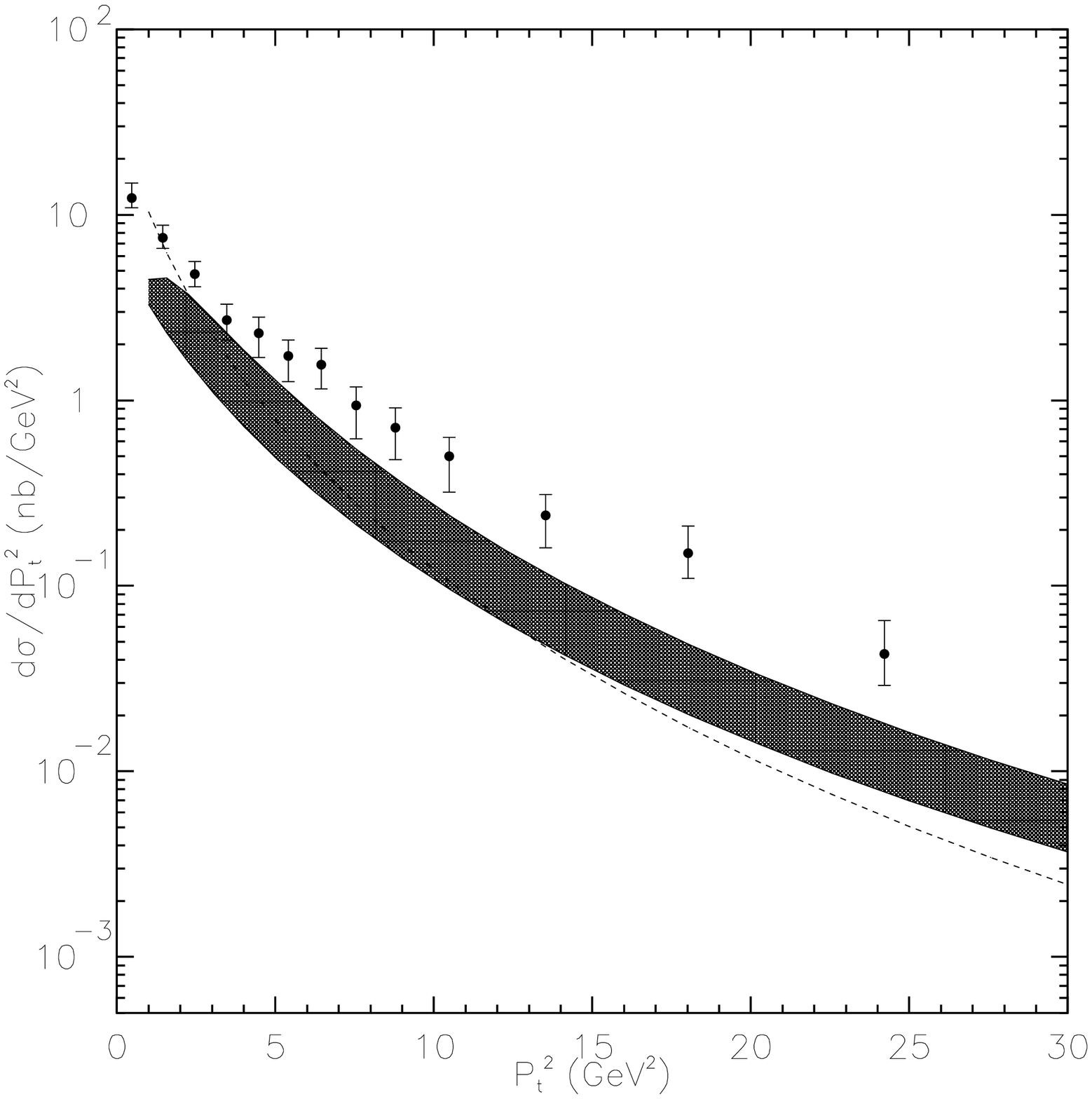}
\caption {\label{fig:zdis} The energy fraction $z$ distribution of $J/\psi$
photoproduction with $P_t>$1GeV (left panel). The transverse momentum $p_t$ distribution 
of $J/\psi$ with $0.4<z<0.9$ (right panel). The dotted lines in each one are the LO results. The
upper bound of the shaded band is obtained with
$\mu_r=\mu_f=\frac{1}{2}\sqrt{(2m_c)^2+p_t^2}$ and $m_c$=1.4GeV, and
the lower one with $\mu_r=\mu_f=2\sqrt{(2m_c)^2+p_t^2}$ and
$m_c$=1.6GeV. The experimental data (filled circles with bars) are
taken from Ref. \cite{Chekanov:2002at}.}}
\end{figure*}

\begin{figure*}
\center{
\includegraphics*[scale=0.31]{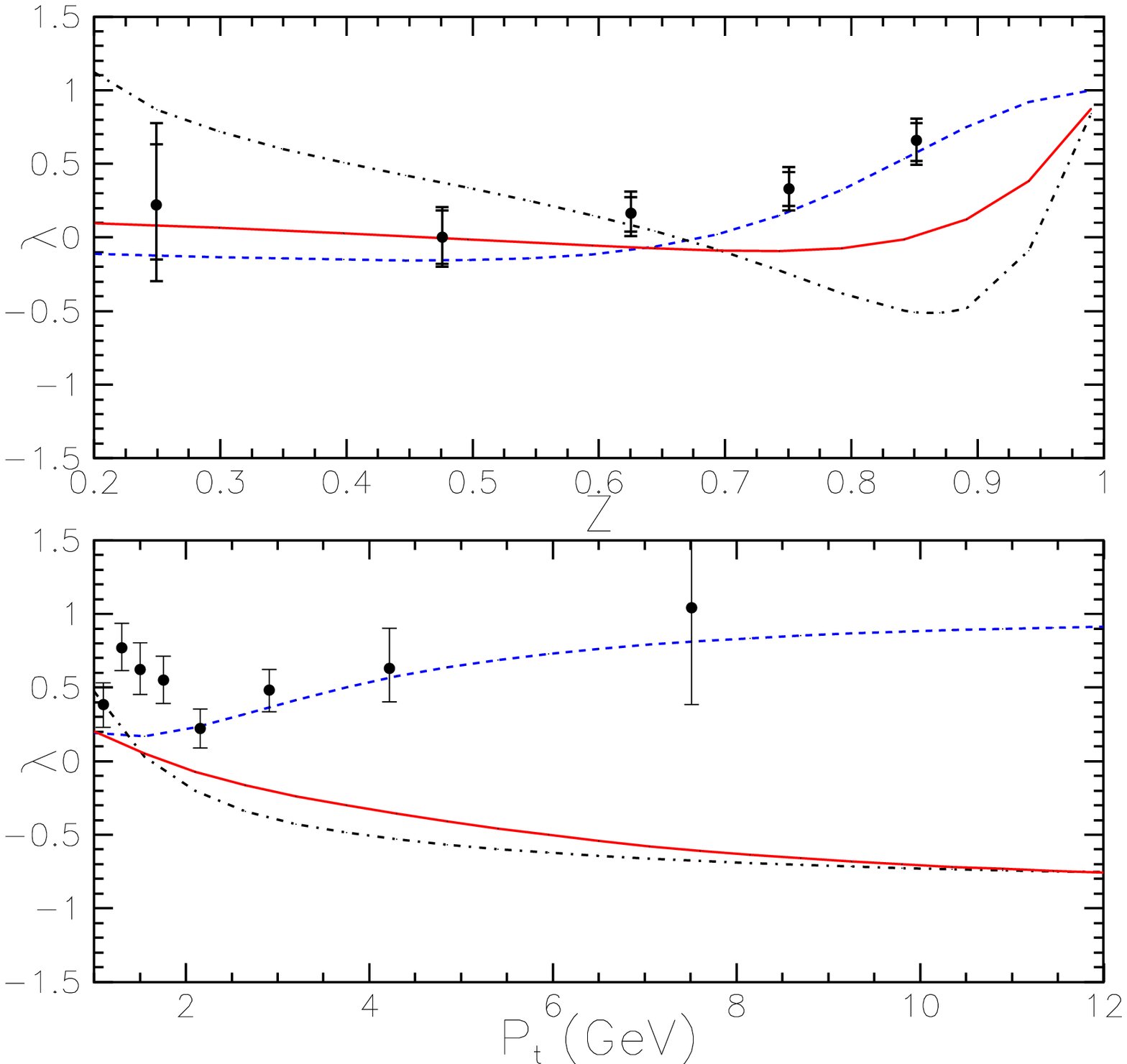}
\includegraphics*[scale=0.31]{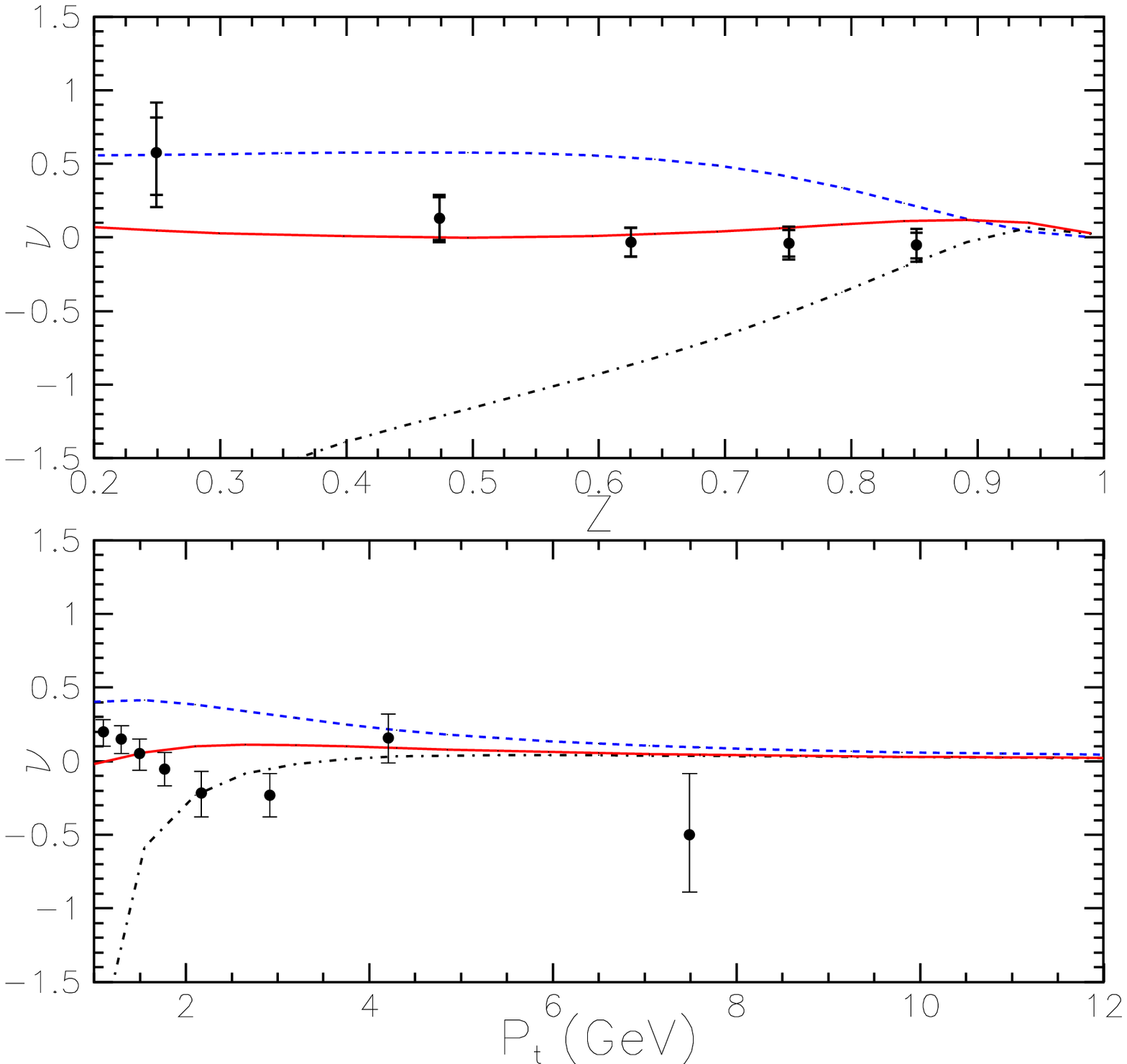}
\includegraphics*[scale=0.31]{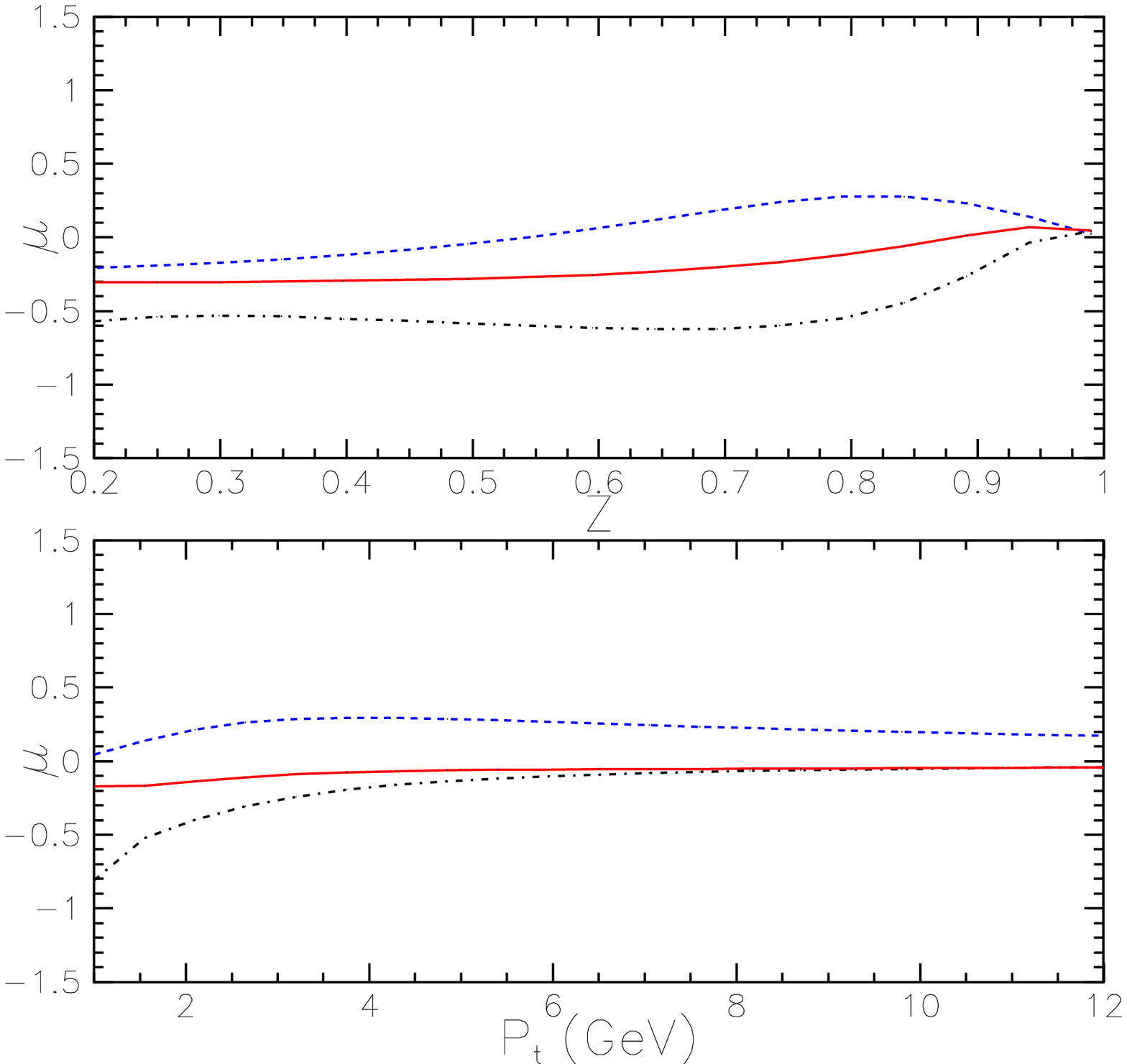}
\caption {\label{fig:lambda} The energy fraction $z$ distributions of polarization
parameters $\lambda,\nu,$ and $\mu$ with $p_t>1$GeV, and 
transverse momentum $p_t$ distributions with $0.4<z<1$. 
Dashed lines are the LO results and the dot-dashed lines are the NLO results.
Solid lines present the results with $\mu_r=\mu_f=8m_c$.
The experimental data (filled circles with bars) are taken from Ref.
\cite{Jungst:2008ip}.}}
\end{figure*}

To calculate the $J/\psi$ photoproduction at NLO, the following relevant
processes \bea
&\gamma+g \to J/\psi +g, \label{eqn:v}  \\
&\gamma+g \to J/\psi + g +g, \label{eqn:r1}  \\
&\gamma+g \to J/\psi + q+\bar{q}, \label{eqn:r2}  \\
&\gamma+ q(\bar{q}) \to J/\psi + g +q(\bar{q}) \label{eqn:r3}. \eea
need to be considered. For process
(\ref{eqn:v}), there are 6 Feynman diagrams at LO and 111 at NLO, 
and there are the ultraviolet(UV), infrared(IR) 
and Coulomb singularities. By using 
dimensional regularization and introducing a small
relative velocity between the quark and antiquark, we can separate the
singularities out. The UV and Coulomb singularities are absorbed into
the redefinition of mass, coupling constant, fields of quark and
gluons and wave function of $J/\psi$ by renormalization, for which
the same renormalization scheme as in Ref.~\cite{Gong:2008sn} is
applied. By using the phase space slicing
method\cite{Harris:2001sx}, we separate out the IR singularities in
the real processes (\ref{eqn:r1}) (\ref{eqn:r2}) (\ref{eqn:r3}). 
Finally to add all the contributions from the real
and virtual parts together, a finite result, which is
free from all the singularities, is obtained.

As the main purpose of this paper, we extract the information of the
$J/\psi$ polarization through the angular distribution of the lepton
$l^+$ in $J/\psi\rightarrow l^+ l^-$ decay. As in Ref.\cite{Beneke:1998re}, the decay angular
distribution of the outgoing $l^+$ can be parametrized in the
$J/\psi$ rest frame as 
\bea \frac{d\sigma}{d\Omega dy} & \propto & 1+
\lambda(y)\cos^2\theta +\mu(y)\sin2\theta \cos\phi \nonumber \\
 &&+\frac{\nu(y)}{2}\sin^2\theta\cos2\phi,\label{eqn:para0}
\eea 
where y stands for a suitable variable (such as transverse
momentum $p_t$, energy fraction $z$, etc.). $\theta$ and
$\phi$ are the polar and azimuthal angles of the outgoing $l^+$, 
respectively. The polarization parameters, $\lambda$, $\mu$, 
and $\nu$, are related to the density matrix
of $J/\psi$ production as 
\bea
\lambda(y)=\frac{d\sigma_{11}/dy-d\sigma_{00}/dy}{d\sigma_{11}/dy+d\sigma_{00}/dy}, \nonumber \\
\mu(y)=\frac{\sqrt{2}Re d\sigma_{10}/dy}{d\sigma_{11}/dy+d\sigma_{00}/dy}, \nonumber  \\
\nu(y)=\frac{2d\sigma_{1-1}/dy}{d\sigma_{11}/dy+d\sigma_{00}/dy}
\label{eqn:para}. 
\eea 
Here $d\sigma_{\lambda\lambda'}/dy$ are the
differential density matrix elements and defined as 
\bea
\frac{d\sigma_{\lambda\lambda'}}{dy}&=&\frac{1}{F}\int
\prod_{i=1}^{n}\frac{d^3p_i}{2E_i}
\delta^4(p_a+p_b-\sum^n_{i=1} p_i) \nonumber \\
&&\delta(y-y(p_i))M(\lambda)M^*(\lambda'), 
\eea 
where $M(\lambda)$ is the matrix element of 
polarized $J/\psi$ production, $\lambda$ and
$\lambda'$ stand for the polarization,
F includes the flux factor and spin average factor,  and $p_i$ is the
momentum of corresponding particles.
In the calculation, the polarization of $\jpsi$ must be explicitly retained
and the treatment for it is the same as in Ref.~\cite{Gong:2008sn}.
Obviously these polarization parameters depend on the coordinate system choice.
In the following, we
calculate them in the target frame with the Z axis defined as the
inverse direction of the initial proton and the polarization vectors of
$J/\psi$ defined in the appendix of Ref.~\cite{Beneke:1998re}. 
We also compute the polarization parameters of
$J/\psi$ in the helicity-base frame where the Z axis is defined as
the $J/\psi$ flight direction in the laboratory frame, and the angular distribution 
of $l^+$ is parametrized as 
\bea
\frac{d\sigma}{d\cos \theta dy} & \propto & 1+
\alpha(y)\cos^2\theta.
\label{eqn:heli} 
\eea 
Here the parameter $\alpha$ is related to the
polarized cross sections of $J/\psi$ production by 
\be
\alpha(y)=\frac{{\md\s_T}/{\md y}-2 {\md\s_L}/{\md
y}}{{\md\s_T}/{\md y}+2 {\md\s_L}/{\md y}}\,, 
\ee 
where $\s_T$ and
$\s_L$ are the cross sections of transverse and longitudinal
polarized $J/\psi$ respectively. $\a=-1$ corresponds to fully
longitudinal polarization and $\a=1$ to fully transverse
polarization.

To replace the polarization vectors for photon or gluons by their
corresponding momentum in the numerical calculation, the gauge invariance is
obviously observed.
Since the two phase space cutoffs are chosen to handle the IR singularities of the real processes,
we numerically check the independence of the results on the cutoffs.

In the numerical calculation, we use $\alpha$=1/137, $m_c$=1.5GeV, and
$M_{J/\psi}=2m_c$. The Cteq6L1 and Cteq6M\cite{Pumplin:2002vw} are used in the calculations 
at LO and NLO respectively, with the corresponding $\alpha_s$ running formula in the Cteq6 
being used. In
Ref.~\cite{Kramer:1995nb} a fixed value of $\alpha_s$ is taken for
numerical calculations. Whereas the running coupling constant is chosen in most literatures, here we use
the running $\alpha_s$ in all our calculations. The renormalization
scale $\mu_r$ and the factorization scale $\mu_f$ are set as
$\mu_r=\mu_f=\sqrt{(2m_c)^2+p_t^2}$. The value for $J/\psi$ wave function at the origin
is extracted from the leptonic decay width with the NLO formula,
the value $\Gamma_{J/\psi \to ee}=$5.55keV, and
$\alpha_s(M_{J/\psi})=$0.26. 
The typical HERA photon-proton center-of-mass energy $\sqrt{s_{\gamma p}}=100$GeV is chosen. 
In addition, the experimental cut conditions $0.4<z<0.9$ and $p_t>1$ for the $p_t$ and $z$
distributions on $J/\psi$ production are applied respectively, and 
$0.4<z<1$ and $p_t>1$ are applied for the $J/\psi$ polarization. 
Furthermore, all the differential cross sections are
calculated directly in numerical calculation by using the
corresponding analytic phase space treatment just like that used in
Ref.~\cite{Gong:2008sn}.

The final results are presented in the figures. From 
Fig.\ref{fig:zdis}, we can see that the results for $p_t$ and $z$ distributions of 
$J/\psi$ production at NLO moderate the
discrepancies between theoretical predictions and experimental
measurements substantially. The $p_t$ and $z$ distributions of the polarization parameters in
Eq.(\ref{eqn:para0}) is presented in
Fig.~\ref{fig:lambda}. In the presentation, we make a cutoff for the
region $z\leq 0.2$, because in this region the denominator of
Eq.~\ref{eqn:para} crosses zero at a certain point so that the values
of the parameters change dramatically in the neighborhood of the
point. It means that the perturbation expansion is
very bad and the obtained result cannot be trusted in this region.
Comparing with the results at LO, the NLO QCD corrections greatly change 
the distributions. The parameters $\lambda$ obtained at LO and
NLO are very different at the large and small $z$ regions (there is a
cross at about $z \approx 0.67$). The $z$
distribution of parameter $\nu$ changes drastically in the
small and intermediate regions, and $\mu$ also changes in the 
intermediate region of $z$. As for the $p_t$ distribution of the
parameters, from the figure one can see that the NLO QCD
correction changes that of the $\lambda$ parameter from positive
values to negative ones and makes it tend to -0.8 as $p_t$
increases. For the $\nu$ parameter, there is little difference
between the results of LO and NLO in the large $p_t$ region. 
The influence of the NLO correction on the $p_t$ distribution of $\mu$ 
is also quite large. For comparison,
the available experimental data at HERA~\cite{Jungst:2008ip} are plotted
in the figures. It is clearly shown that both LO and NLO results
do not fit the polarization measurement, and the NLO results are even worse. 
In these figures, the z and $p_t$ distributions of 
$\lambda,\mu,$ and $\nu$ with $\mu_r=\mu_f=8m_c$ at NLO are also presented.

\begin{figure}
\center{
\includegraphics*[scale=0.35]{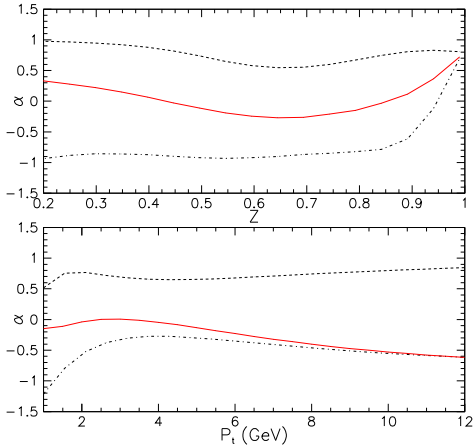}
\caption {\label{fig:helicity}The polarization parameter $\alpha$
distributions as functions of $P_t$ and $z$ in the helicity basis.
Dashed lines are the results at LO and the dot-dashed lines are
the ones at NLO. The results with $\mu_r=\mu_f=8m_c$ are presented by the solid lines.}}
\end{figure}

The $p_t$ and $z$ distributions of $\alpha$ in the helicity basis are 
presented in Fig.~\ref{fig:helicity}. We also make a cutoff for the
region $z\leq 0.2$, due to the same reason as that in the case of 
the target frame. 
It is similar to the situation in the
$J/\psi$ hadroproduction: NLO QCD corrections change the $p_t$
distribution of $\alpha$ from positive values to negative ones. The
$z$ distribution of $\alpha$ changes similar to that
on the $p_t$ distribution except the endpoint region near $z=1$. We
also present the results with $\mu_r=\mu_f=8m_c$.

In summary, we have investigated the photoproduction of $J/\psi$ at QCD 
NLO at HERA. The results show that the NLO QCD corrections cannot 
give a very good description on the $p_t$ and 
$z$ distributions. Furthermore, for the $J/\psi$ polarization, 
the NLO QCD corrections make a drastic change on certain
polarization parameters, but they cannot give a satisfied prescription for
the available experimental data. In contrast to the LO calculation for polarization, 
the NLO results are even worse. 
There are two conclusions on the results at QCD NLO. One is 
that the theoretical uncertainties for the z distributions of the $J/\psi$ polarization parameters 
$\lambda,\mu,\nu,$ and $\alpha$ on different choices of the renormalization and factorization 
scales are too large to give reasonable predictions to compare with the experimental measurement.  
Another is that the theoretical uncertainties for the $p_t$ distributions of  
$\lambda,\mu,\nu,$ and $\alpha$
on different choices of the scales are small when $p_t>3GeV$ and 
the obtained $p_t$ distributions cannot describe the experimental data even just in this region. 
Therefore, there is still no satisfactory theoretical description even at the NLO level 
on the $p_t$ or z distributions of $J/\psi$ polarization as well as the $p_t$ or z distribution
of $J/\psi$ productions at HERA. It will be interesting
to know the situation by considering NLO color-octet contributions, or higher
order QCD corrections in a future study. 

While this paper was being prepared, we were informed of the same process
also being considered by Artoisenet et~al.~\cite{Artoisenet:2009xh}. 
Comparing our results in Figs. 1 and 2 with theirs, there are quantitative discrepancies 
between them. But our results on the z distributions of $\lambda$ and $\nu$ 
with $\mu_r=\mu_f=8m_c$ at NLO are consistent with theirs with $\sqrt{s_{\gamma p}}=100GeV$
and $\mu_r=\mu_f=8m_c$\footnote{Private communication with P. Artoisenet and F. Maltoni}. It can be inferred that the discrepancies mainly come from the 
different choices of renormalization and factorization scales. 

We would like to thank Gong Bin for
helpful discussions, M. Kramer for providing information about his work. 
We also thank P. Artoisenet and F. Maltoni for comparison. 
This work was supported by the National Natural
Science Foundation of China (No.~10775141, No.~10547001, and No.~10875155) and
by the Chinese Academy of Sciences under Project No. KJCX3-SYW-N2.

\end{document}